\documentclass{ijuc}
\usepackage[pdftex]{graphics}
\usepackage{algorithm}
\usepackage{algpseudocode}
\usepackage{amsmath}

\usepackage{color}
\definecolor{ForestGreen}{RGB}{162,52,0}
\usepackage{xcolor}
\usepackage{amsmath}
\usepackage{CJK}
\usepackage{url}
\usepackage{array}
\usepackage{cite}

\begin{document}

\title{A multi-domain virtual network embedding algorithm with delay prediction}

\author{Peiying Zhang\inst{1}
\and Xue Pang\inst{1}
\and Yongjing Ni\inst{2,3}
\and Haipeng Yao\inst{4} \email{yaohaipeng@bupt.edu.cn}
\and Xin Li\inst{5}
}

\institute{College of Computer Science and Technology, China University of Petroleum (East China), Qingdao 266580, China.
\and
School of Information Science and Engineering, Yanshan University, Qinhuangdao 066000, China.
\and
School of Information Science and Engineering, Hebei University of Science and Technology, Shijiazhuang 050000, China.
\and
Beijing University of Posts and Telecommunications, Beijing 100876,China.
\and
College of Information Technology and Cyber Security, People's Public Security University of China, Beijing 100038, China.
}

\def\received{Received 15 January 2020; In final form}
\maketitle

\begin{abstract}
Virtual network embedding (VNE) is an crucial part of network virtualization (NV), which aims to map the virtual networks (VNs) to a shared substrate network (SN). With the emergence of various delay-sensitive applications, how to improve the delay performance of the system has become a hot topic in academic circles. Based on extensive research, we proposed a multi-domain virtual network embedding algorithm based on delay prediction (DP-VNE). Firstly, the candidate physical nodes are selected by estimating the delay of virtual requests, then particle swarm optimization (PSO) algorithm is used to optimize the mapping process, so as to reduce the delay of the system. The simulation results show that compared with the other three advanced algorithms, the proposed algorithm can significantly reduce the system delay while keeping other indicators unaffected.
\end{abstract}


\section{Introduction}

In modern society, all activities cannot be separated from the participation of Internet, the whole world has become an interconnected entirety. With the improvement of living standard, people's demands is higher and higher, in order to meet people's demands for the function and performance of the Internet, Internet-based services are developing rapidly, however, the rigidity of traditional networks is increasingly serious. Therefore, the network virtualization (NV) has emerged\cite{DBLP:journals/access/ZhangYL16}.

NV allows multiple parallel and isolated VNs share the same resources of one SN, a SN may contain one or more domains\cite{Li2016Multi}. For example, some modern medical technologies require low delay performance, while some backup projects require high bandwidth performance. Therefore, sharing the SN resources among multiple VNs can increase the network elasticity, each VN can deploy different protocols to meet different needs so that achieve functional diversification, making it more flexible and reasonable to share the resources, thus effectively solving the rigidity problem of traditional network architecture\cite{DBLP:journals/comcom/Zhang18}. Participants in traditional VNE include infrastructure providers (InPs) and service providers (SPs), the InPs are responsible for deploying the devices of SN and providing physical resources, while the SPs are responsible for creating and managing the VNs. The purpose of VNE is to find an optimal mapping result based on satisfying the resource constraints\cite{Peiying2018Topology}.

\begin{figure}
\begin{center}
\scalebox{0.3}{\includegraphics{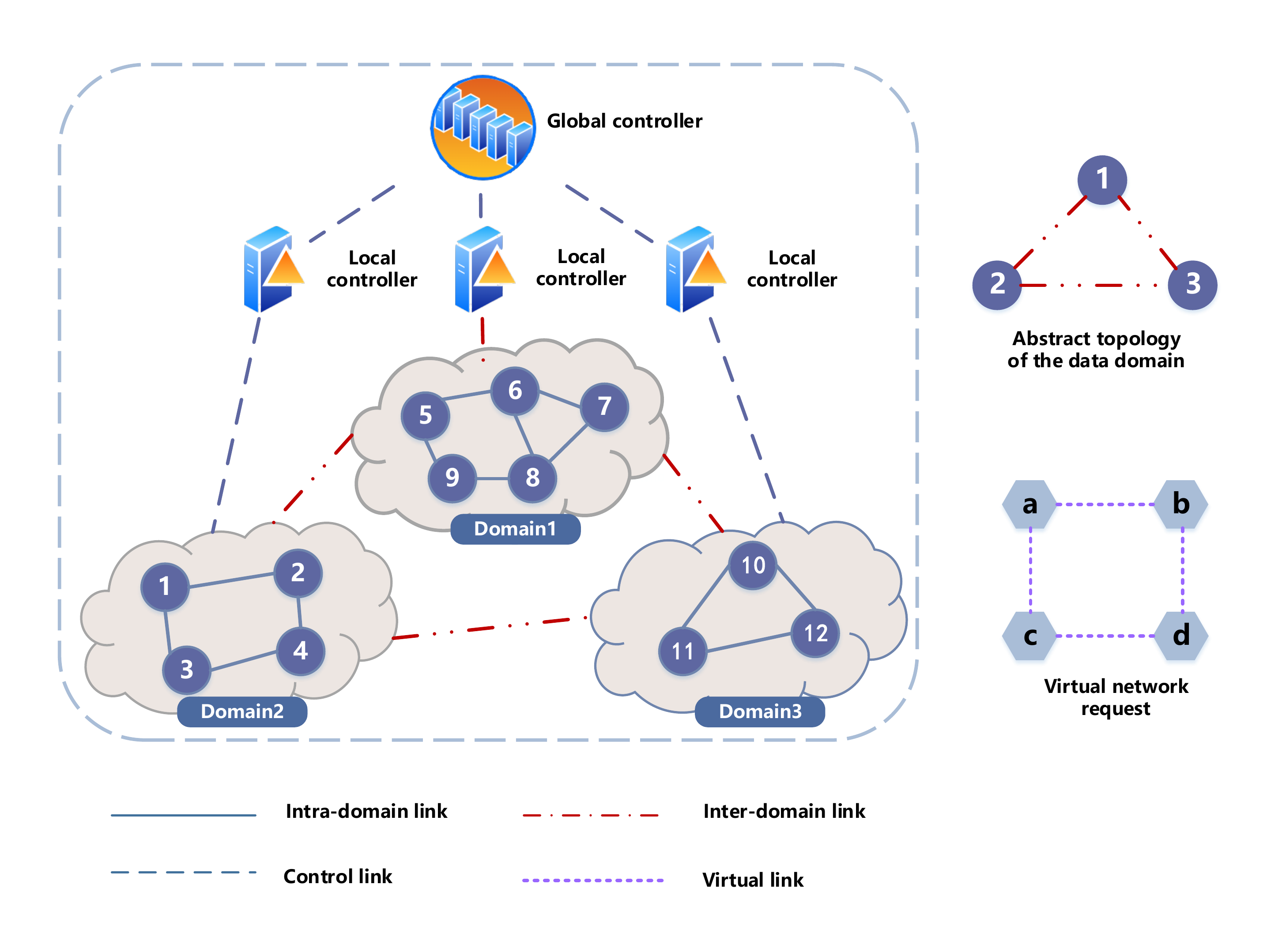}}
\end{center}
\caption{The diagram of SDN architecture.}
\label{fig-eg1}
\end{figure}

In traditional VNE process, each SP has to constantly discuss with multiple InPs when establishing a cross-domain virtual network request, which is very troublesome. While in multi-domain software defined networks (SDN), the separation of control plane and data plane is realized, thus effectively improving the mapping efficiency of VNs. In this paper, we use multi-domain SDN architecture as the mapping architecture. As shown in FIGURE \ref{fig-eg1}, the multi-domain SDN VNE architecture model consists of four parts:

(1) Service Providers (SPs): SPs are responsible for sending VN requests to the global controller of SN.

(2) Global Controllers: Global controllers are responsible for receiving VN requests from SPs and sending them down to local controllers, and then mapping the VNs based on the information uploaded by the local controllers.

(3) Local Controllers: A local controller manages a data domain, collects information that needs to be uploaded to the global controller, and provides VN resources for VN requests.

(4) Data Domains: Data domain is the data plane of SDN. A data domain consists of several switches managed by a local controller.

Among numerous web-based services, more and more applications are delay-sensitive, low time delay plays an increasingly important role in modern medical care, intelligent transportation, modern army and other fields. In the field of modern medicine, low time delay plays an important role in remote surgery, doctors can accurately grasp the process of surgery and the condition of patients, so they can provide accurate guidance. In the field of intelligent transportation, driverless technology has a very high requirement for low delay. If sensors in cars can quickly receive the information provided by the Internet of vehicles, the braking distance can be shortened in an emergency, thus ensuring the distance between vehicles and pedestrians. In the modern military field, the army's network system has a particularly high demand for time delay, in the rapidly changing battlefield, it is difficult to make accurate judgment and master the initiative of communication if we cannot timely grasp the trend of the battlefield. In addition to these areas, low delay also affects every aspect of our life. Online games need lower delay, in some games with fierce competition, if the transmission of operation is delayed, it is likely to face the risk of losing the game. In the modern office environment, with the emergence of more and more online meetings, smooth pictures can improve people's experience, which is accomplished by low delay. When browsing the web, we all hope to query the content quickly, if time delay is too high, the user experience will be greatly decreased, so people also devoted more energy to study how to reduce the time delay.

In this paper, in order to optimize the delay in the process of multi-domain virtual network embedding, we proposed a multi-domain virtual network embedding algorithm with delay prediction. The main contributions of this algorithm are as follows:

(1) A method to select candidate physical nodes by predicting the delay is proposed. This algorithm realizes the selection of candidate physical nodes by predicting the delay of VNs, which is the first step to reduce the mapping delay.

(2) Candidate node information is collected and uploaded to the global controller by local controllers based on SDN architecture, the deployment of SDN architecture further improves the mapping efficiency.

(3) After the global controller receives the candidate node information, it uses the PSO algorithm to generate the VNE result, and sends it to the local controller to complete the mapping. The use of PSO algorithm can better select the optimal solution.

The reminder of this paper is organized as follows. Section 2 reviews the existing methods for VNE. Section 3 introduces the network model and problem statement. In Section 4, we describe our proposed method DP-VNE in detail. Section 5 introduces several key algorithms. The performance of our method and other methods is evaluated in Section 6. Section 7 concludes this paper.

\section{Related Works}
In this section, we will review the existing literature on virtual network embedding algorithms across multiple domains. These existing algorithms can be roughly partitioned into two categories including the distributed algorithms and the centralized algorithms.

\subsection{The Distributed VNE Algorithms}
At present, many networks are deployed in multi-domain environments, most studies assume that the SNs run smoothly and without interruption, but in fact, most of the existing research only stays in the level of single domain, the SNs cannot guarantee normal services all the time, so the traditional single-domain mapping algorithm cannot meet the needs of users. Therefore, the authors of literature \cite{Xiancui2017} proposed a multi-domain survivable virtual network mapping algorithm, this algorithm can map the virtual network to different domains, provide backup resources for virtual links, and improve the survivability of the special network effectively. Recent advances in software-defined mobile networks (SDMNs), in-network caching, and moving edge computing (MEC) could have a significant impact on services in the next generation of mobile networks \cite{DBLP:journals/twc/LiangHYZ18}.

Block-chain applications in wireless mobile networks have been hampered, which puts a high demand on the computing power and storage availability of mobile devices. To solve the above problems, the authors of literature \cite{DBLP:journals/tvt/LiuYTLS18} proposed a new mobile edge computing (MEC) wireless block-chain framework. The authors of \cite{YaoRDAM} proposed a reinforcement learning based dynamic attribute matrix representation for virtual network embedding (RDAM). In literature \cite{M2M}, the authors introduced mobile edge computing (MEC) into the virtual cellular network with machine-to-machine (M2M) communication, which can reduce energy consumption and improve computing power. In addition, to facilitate the integration of the network architecture, the SDN was introduced to handle the various protocols and standards in the network. This work has effectively improved the performance of the system. At the same time, non-trivial enhancement learning is used to train the virtual network embedding algorithm. Due to the selfishness of different InPs, they are reluctant to disclose the specific topology information of networks. Therefore, the authors of literature \cite{Wangxiaofei2017} proposed a effective two-step strategy of multi-domain virtual network embedding in 5G network slicing, which can dynamically manage the resource information of different domains. In literature \cite{Yuchen}, the authors proposed a new information-centric heterogeneous network framework, which gives nodes the ability to store and compute, thus improving data throughput. In addition, this article described the virtual resource allocation strategy as a joint optimization problem that considers not only the benefits of virtualization but also the benefits of caching and computing. Finally, a distributed algorithm based on alternating direction multiplier is used to solve the problem, which can greatly reduce the computational complexity and signaling overhead. The authors of \cite{Peiying2019Security} proposed a security aware virtual network embedding algorithm using information entropy TOPSIS to solve the security problems in the process of virtual network embedding. The authors of \cite{DBLP:journals/ppna/ZhangYLL19} proposed a virtual network embedding algorithm based on modified genetic algorithm to increased the acceptance ratio of VNRs and the long-term average revenue of Infrastructures (InPs). The authors of literature \cite{DBLP:conf/sigcomm/WerlePHLZBM11} proposed a virtual network provisioning platform across multiple domains, which could decompose the virtual network into multiple parts to avoid the impact of limited disclosure of information on resource discovery and distribution. The platform allows a new virtual network request to be assigned to a different domain and to obtain the resources of each domain. In addition, a signaling protocol is designed to realize resource reservation for the virtual link setup between domains, so that the virtual link with high quality service can be provided. The authors of literature \cite{DBLP:journals/osn/YuWDAL14} proposed a low-cost VN mapping framework with joint intra-domain and inter-domain mapping. This framework points out that the virtual network mapping is completed by the mapping manager, which is the agent between InPs and SPs. The existing cross domain virtual network mapping algorithm is mainly to decompose a virtual network into multiple subnetworks. There is little research on the connection relationship between inter-domain links and intra-domains links. The authors of literature \cite{DBLP:conf/softcom/2016} proposed a multi-domain link mapping algorithm, which is an Internet-based algorithm with good deployability. Block-chain is widely regarded as a promising solution, which can solve the security and efficiency problems of massive industrial Internet of things (IIoT) data.

In order to meet the requirements of high throughput, literature \cite{DBLP:journals/tii/LiuYTLS19}\cite{DBLP:journals/fgcs/QiuCYXYZ19} proposed a new deep reinforcement learning (DRL) based block-chain supporting IIoT system performance optimization framework. The authors of \cite{DBLP:journals/cn/HouidiLBZ11} discussed the configuration of virtual resources in future networks based on infrastructure service principles. A precise and heuristic optimization algorithm for virtual network provisioning involving multiple infrastructure providers is proposed. It effectively reduces the embedding cost of infrastructure providers and improves the acceptance rate of requests. In the Internet of things infrastructure based on network function virtualization (NFV), the service function chain (SFC) is an ordered combination of interconnected virtual network functions (VNFs) based on the application logic of the Internet of things. In literature \cite{NFV-Enabled}, the authors decomposed complex VNFs into smaller virtual network functional components (VNFCs) to make more effective decisions, and proposed a deep enhanced learning (DRL) scheme based on experience playback and target network, which can effectively handle complex and dynamic SFC embedded Internet of things scenarios. And in literature \cite{Xiaoyuan}, a deep reinforcement learning (DRL)-based scheme with experience replay and target network is proposed as a solution. There is no suitable solution to the multi-domain virtual network embedding problem, because most studies assume that the SN can run smoothly without any interruption. However, due to external reasons, the SN cannot guarantee the normal provision of network services. In order to solve the above problems, the authors of \cite{DBLP:journals/scn/XiaoZZ17} proposed a multi-domain survivable virtual network mapping algorithm (IntD-GRC-SVNE), which maps virtual communication network to different domain networks, provides backup resources for virtual links, and improves the survivability of special networks. Mobile edge computing (MEC) is an important component of the next generation network, in order to enhance processing power and provide low-latency computing services for the Internet of things (IoT), literature \cite{DBLP:journals/iotj/QiuYYJXZ19} proposed a resource allocation strategy to maximize the available processing power of the IoT network.

In most multi-domain virtual network embedding algorithms, traditional network architecture are used, and participants are InPs and SPs. However, our algorithm is similar to some algorithms, we use the SDN architecture to separate the data plane and control plane, which will undoubtedly improve the efficiency of the algorithm.

\subsection{The Centralized VNE Algorithms}
Aiming at the problem of virtual network embedding in multi-domain network environment, the authors of \cite{Peng2015A} proposed a multi-domain virtual network embedding algorithm based on minimum cost. First, according to the constraints of the virtual network, the feasible substrate node set of each virtual node is calculated. Then the minimum weight routing algorithm is used to calculate the feasible substrate path set for each virtual link. Finally, based on the minimum spanning tree algorithm of Kruskal, the lowest option value of the feasible set of substrate paths is sequentially selected. In literature\cite{DBLP:journals/tvt/QiuYYXZ19}\cite{DBLP:journals/comsur/XieYHXLWL19}, the authors combined deep learning and blockchain with network, transforms the joint resource allocation problem into a joint optimization problem, and uses deep learning to solve the problem. The authors of literature\cite{DBLP:journals/iotj/ZhangYL18} proposed a VN embedding model based on 3D resource constraints such as computing, network and storage, and designed two heuristic algorithms as the baseline algorithm to deal with the VN embedding problem. The authors of \cite{YuCost} proposed a cost efficient virtual network mapping across multiple domains with joint intra-domain and inter- domain mapping. In this framework, a proxy between SPs and InPs is completed, called the mapping manager. The candidate selection problem is transformed into a mixed integer linear programming (MILP) problem. A key phrase embedded in multi-domain virtual networks is VN partitioning, which divides a VN into multiple physical domains.

Since the multi-domain VNE problem is np-hard, in order to improve the efficiency of VN partition, the authors of \cite{Guo2015Particle} proposed a particle swarm optimization based multi-domain virtual network embedding algorithm(VNP-PSO). VNP-PSO algorithm generates the approximate optimal solution of VN partition through the evolution of particles. Simulation results show that this algorithm improves the efficiency of VN partition and reduces the embedding cost of multi-domain VNE. The authors of \cite{ZhangMulti}\cite{DBLP:journals/access/ZhangYQL18} proposed the idea of using node resource measurement to solve the virtual network embedding problem, which effectively improved the performance indexes in the process of virtual network embedding. In order to solve the np-hard problem, the authors of \cite{Shahin2015Memetic} proposed a memetic multi-objective particle swarm optimization-based energy-aware virtual network embedding, which can accelerate the convergence speed of the algorithm and improve the quality of the solution. Mobile edge computing (MEC) is an important part of the next generation network, which can improve the available processing power of the Internet of things (IoT) and provide low-latency computing services for the Internet of things (IoT). The authors of \cite{DBLP:journals/iotj/QinCZCYW19} investigated a resource allocation strategy to maximize the available processing power of the MEC Internet of things network. The NV has been widely accepted as a technology for future networks. The authors of \cite{Cao2018A} proposed a novel optimal mapping algorithm with less computational complexity for virtual network embedding. This algorithm constructs a subset of candidate substrate nodes and a subset of candidate substrate links before embedding the virtual network, thus greatly reduces the execution time of the mapping without causing a performance penalty.

Traditional heuristic mapping algorithms use static processes leading to suboptimal ordering and embedding decisions. The authors of \cite{DBLP:journals/ijon/YaoCLZW18} introduced the method of reinforcement learning into virtual network embedding. The authors of \cite{Yu2016Virtual} focused on a multi-domain virtual network embedded in a heterogeneous 5G network infrastructure that facilitates resource sharing of different functional requirements from end users. They proposed a virtual 5G network embedding in a heterogeneous and multi-domain network infrastructure, which can accurately divide 5G embedding requirements to match different user access densities. The algorithm performs better in average blocking rate, routing delay and wireless or wired resource utilization. The authors of \cite{DBLP:conf/wpmc/CuiGLL13} proposed a virtual network embedding algorithm based on virtual topology connection feature and introduce a node connectivity degree to measure the priority of candidate physical nodes. The connection degree of nodes based on virtual topology features is used to select nodes that are close to each other, thus making the algorithm more efficient. In literature\cite{DBLP:journals/cm/HeYZLY17}, the authors combined the moving edge computing with the network, which effectively improves the performance of the network.

PSO algorithm is widely used in many schemes, similar to literature \cite{gengruiwen}, our algorithm also uses PSO algorithm to generate mapping results. The difference is that we define a formula to predict the node delay, select the best candidate node by predicting the node delay, and define the target formula by taking the delay as the optimization objective.

\section{Network Model and Problem Statement}
\subsection{Virtual Network Model}

In general, we use a weighted undirected graph $G_{V}=\left \{ N_{V},L_{V} \right \}$ to represent a virtual network, therein, $N_{V}=\left ( n_{V}^{i},i=1,2,3,\cdots \right )$ represents the set of all virtual nodes, the CPU resource demand of each virtual node can be expressed as $CPU\left ( n_{V}^{i} \right )$, and the candidate domains of each virtual node are represented by $CD\left ( n_{V}^{i} \right )$ , $L_{V}=\left \{ l_{V}^{k},l_{V}^{k}=\left ( n_{V}^{i},n_{V}^{j} \right ),k=1,2,3\cdots \right \}$ represents the set of all virtual links, $BW\left ( l_{V}^{k} \right )$ is used to represent the bandwidth resource demand of virtual link $l_{V}^{k}$.

\begin{figure}
\begin{center}
\scalebox{0.4}{\includegraphics{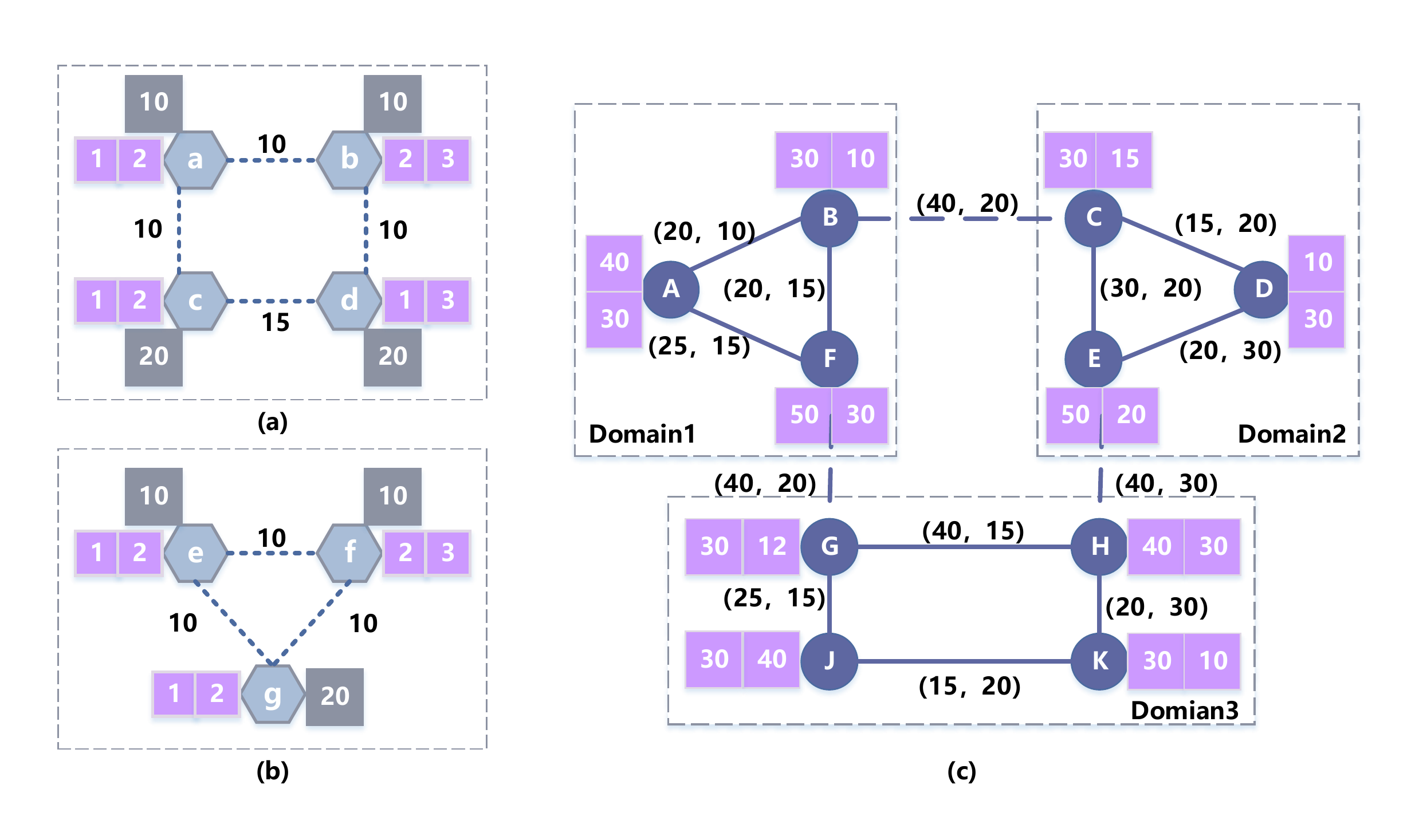}}
\end{center}
\caption{The diagram of virtual network and substrate network.}
\label{fig-eg2}
\end{figure}

An example of virtual network is shown in FIGURE \ref{fig-eg2}(a)(b). There are two virtual networks in the figure. Virtual network (a) contains 4 virtual nodes \{$a$, $b$, $c$, $d$\}, the virtual nodes are represented by hexagon, the CPU resource demand of each virtual node is represented by the number in the square, the number in two adjacent squares next to the virtual node represents the candidate domains of the virtual node, each virtual node can have two candidate domains, at the same time, the virtual network also includes 4 virtual links \{$a$-$b$,$b$-$c$,$c$-$d$,$d$-$a$\}, the virtual links are represented by dotted lines, and each virtual link has its corresponding bandwidth resource demand, represented by the numbers beside the dotted lines.

\subsection{Substrate Network Model}
The substrate network can also be represented by a weighted undirected graph $G_{S}=\left \{ N_{S}, L_{S}\right \}$, therein, $N_{S}=\left ( n_{S}^{i},i=1,2,3\cdots \right )$ represents the set of all physical nodes, the available CPU capacity of physical nodes are represented by $CPU\left ( n_{S}^{i} \right )$, and the delay of each physical node is represented by $delay\left ( n_{S}^{i} \right )$, $L_{S}=\left \{ l_{S}^{k},l_{S}^{k}=\left ( n_{S}^{i},n_{S}^{j} \right ),k=1,2,3\cdots \right \}$ represents the set of all physical links, $BW\left ( l_{S}^{k} \right )$ is used to represent the bandwidth capacity of physical link $l_{S}^{k}$, the delay of each physical link is represented by $delay\left ( l_{S}^{k} \right )$.

The substrate network is shown in FIGURE \ref{fig-eg2}(c), this substrate network is divided into three different domains, including 10 physical nodes \{ $A$, $B$, $C$, $D$, $E$, $F$, $G$, $H$, $J$, $K$ \}, the physical nodes are represented by circles, where, $B$,$C$,$F$ and $E$ are boundary nodes. Each physical node has its available CPU capacity, represented by the first square next to the node, and each physical node has its delay, represented by the second square next to the node. At the same time, the substrate network also includes eight physical links \{$A$-$B$, $A$-$F$, $B$-$F$, $C$-$D$, $C$-$E$, $E$-$D$, $G$-$H$, $H$-$K$, $J$-$K$, $G$-$J$, $B$-$C$, $F$-$G$, $E$-$H$ \}, where, \{ $A$-$B$, $A$-$F$, $B$-$F$, $C$-$D$, $C$-$E$, $E$-$D$, $G$-$H$, $H$-$K$, $J$-$K$, $G$-$J$ \} are intra-domain links, represented by solid lines, and \{ $B$-$C$, $F$-$G$, $E$-$H$ \} are inter-domain links, represented by dashed lines. The parentheses next to the physical link have two parameters, the first representing the available bandwidth capacity of the physical link, and the second representing the delay of the physical link.

\subsection{Virtual Network Embedding Problem Description }
VNE is the process of mapping virtual networks to the substrate network, which can be represented by $VNE:G_{V}\rightarrow G_{S}$. The mapping process can be divided into two steps:

(1) Node mapping: Node mapping is the first step of VNE. Only when the candidate physical node meets the resource requirements and specific constraints of the virtual node can the virtual node be mapped to the corresponding physical node.
\begin{equation}
\label{eq:vnr}
CPU(n_{S}^{i})\geq CPU(n_{V}^{j}) .
\end{equation}

Formula (1) indicates that the available CPU resources of the physical node must be greater than or equal to the CPU resources required by the virtual node. In addition, each physical node can only be mapped once, that is, the physical node and the virtual node need to correspond one to one.

(2) Link mapping: The node mapping is followed by the link mapping, that is, the virtual link is mapped to the corresponding physical link. During the mapping process, the following conditions need to be met:
\begin{equation}
\label{eq:vnr}
\sum BW(n_{S}^{m},n_{S}^{n})\geq \sum BW(n_{V}^{u},n_{V}^{v}) ,
\end{equation}

\begin{equation}
\label{eq:vnr}
BW(n_{S}^{m},n_{S}^{n})\geq BW(n_{V}^{u},n_{V}^{v}) .
\end{equation}

Formula (2) indicates that in the link mapping process, the total available bandwidth resources of physical links must be greater than or equal to the total required bandwidth resources of virtual links. Formula (3) indicates that for every one-to-one corresponding link mapping, the available bandwidth resources of physical links must be greater than or equal to the bandwidth resources required by virtual links.

\subsection{Objectives}

The optimization objective function defined in this section is shown in formula (4) :
\[Delay(VN)=\sum_{num(N_{V})}\left [ CPU(n_{V}^{i})*delay(n_{S}^{i}) \right ]+\]
\begin{equation}
\label{eq:vnr}
\sum_{num(L_{V})}\left [ BW(l_{V}^{k})*delay(l_{S}^{k}) \right ].
\end{equation}

Formula (4) is used to measure the delay performance of virtual network, wherein, the first term represents the delay performance of nodes, and the second term represents the delay performance of links. $CPU(n_{V}^{i})$ represents the amount of CPU resources required by the virtual node, $delay(n_{S}^{i})$ represents the delay of the physical node mapped by the virtual node, $BW(l_{V}^{k})$ represents the amount of bandwidth resources required by the virtual link, and $delay(l_{S}^{k})$ represents the delay of the physical link mapped by the virtual link. Our goal is to minimize the optimization objective formula of the final mapping scheme.

\section{Design of Virtual Network Mapping Algorithm}
In the process of multi-domain VNE, the main steps can be described as follows.

\subsection{Divide Virtual Network Requests into Subgraphs}
\begin{figure}
\begin{center}
\scalebox{0.5}{\includegraphics{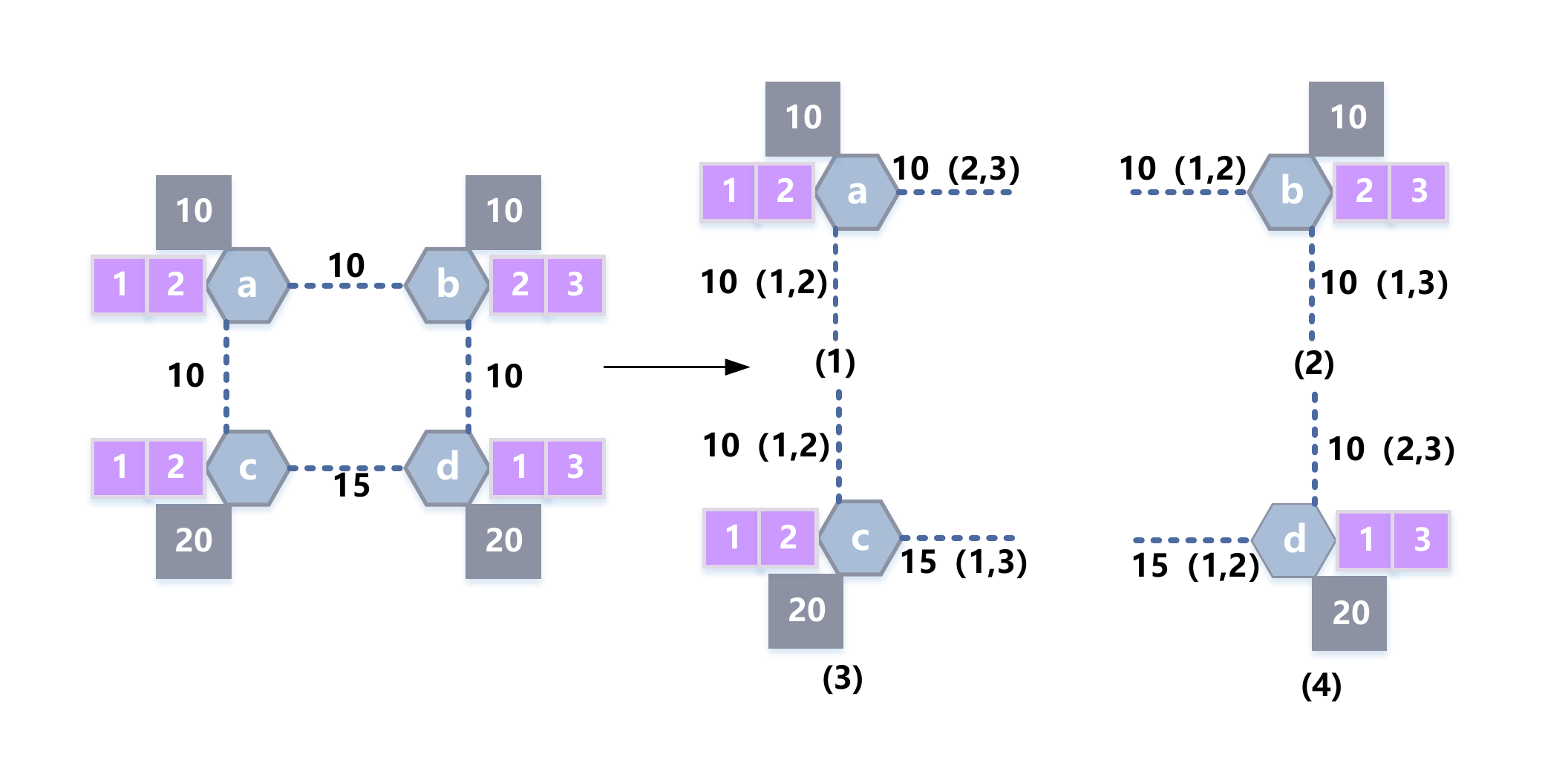}}
\end{center}
\caption{The process of divide virtual network requests into subgraphs.}
\label{fig-eg3}
\end{figure}

The first step of VNE is to divide the received virtual network requests into subgraphs. As shown in FIGURE \ref{fig-eg3}, the virtual network request includes four virtual nodes, so it can be decomposed into four subgraphs. Subgraph 1 includes various resource information of virtual node $a$, the numbers in parentheses on the links at both ends represent candidate domains for the other two nodes connected to $a$, (2,3) indicates that the candidate domains of node $b$ are domain 2 and domain 3, (1,2) indicates that the candidate domains of node $c$ are domain 1 and domain 2.
\subsection{Selection of Candidate Nodes}
After partitioning the virtual request subgraph, for each subgraph, the corresponding candidate physical node is found in the corresponding candidate domain. We use formula (5) to select the candidate physical node. Formula (5) defines a DelayUnit to measure the delay attribute of candidate physical nodes, and finally selects the physical nodes with the smallest DelayUnit as candidate physical nodes.
\[DelayUnit=CPU(n_{V}^{i})*delay(n_{S}^{i})+ \]
\begin{equation}
\label{eq:vnr}
\frac{\sum_{num(L_{V})}\sum_{num(N_{C})}BW(l_{V}^{k})*ldelay(l_{S}^{k})}{num(L_{V})}.
\end{equation}

In formula (5), $N_{C}$ represents the candidate nodes of the virtual nodes, $CPU(n_{V}^{i})$ represents the required CPU resources of the virtual nodes, $delay(n_{S}^{i})$ represents the delay of the candidate physical nodes, $BW(l_{V}^{k})$ represents the bandwidth resource demand of the virtual links, and $ldelay(l_{S}^{k})$ represents the delay measure of physical links, and its value can be divided into two cases. If the candidate physical domain of the two connected virtual nodes are the same, the value is 0. If the candidate physical domain of the two connected virtual nodes is different, its value is the sum of the delay of the mapped physical link.

\subsection{Upload Resource Information}
After the candidate node is selected, the local controllers upload the generated information to the global controller, so that the global controller can generate a pseudo-topology based on the information it receives, and then pre-map it. Among them, the information uploaded by local controllers to the global controller includes the available CPU resources of all candidate physical nodes, the shortest delay path of all candidate physical nodes to boundary nodes, the path delay between candidate nodes, the available CPU resources of boundary nodes, the link delay between boundary nodes, and the available bandwidth resources and delay for all links. The amount of information that needs to be uploaded should not be too large, which would increase the computational burden of the global controller, thus making it impossible to achieve optimization purposes.

\subsection{Pre-mapping of Virtual Network Requests}
After receiving the information of candidate nodes and links, the global controller will generate a pseudo-topology, and make the final decision on the mapping scheme through the generated pseudo-topology using PSO algorithm. PSO algorithm is a random search algorithm based on group cooperation developed by simulating foraging behavior of birds. It is generally considered to be a type of Swarm Intelligence (SI). The updated calculation formula of PSO algorithm's position and velocity is redefined as follows:
\begin{equation}
\label{eq:vnr}
V_{i}^{new}=aV_{i}\oplus b(X_{sp}\ominus X_{i})\oplus c(X_{gp}\ominus X_{i}),
\end{equation}

\begin{equation}
\label{eq:vnr}
X_{i}^{new}=X_{i}\otimes V_{i}^{new}.
\end{equation}

According to the situation in this paper, we defined each parameter according to formula (6) and formula (7) :

(1) Fitness function: In this paper, the objective function formula (4) is used as the fitness function.

(2) Particle position $X_{i}$: $X_{i}=(x_{i}^{1},x_{i}^{2},x_{i}^{3},\cdots x_{i}^{n})$ represents a mapping scheme of virtual network requests, that is, the node number of the global pseudo-topology, and n represents the number of nodes of virtual network requests.

(3) Particle velocity $V_{i}$: $V_{i}=(v_{i}^{1},v_{i}^{2},v_{i}^{3},\cdots v_{i}^{n})$ determines whether the result of virtual network request needs to be changed, where $v_{i}^{k}$ is a binary variable, if the value of $v_{i}^{k}$ is 0, it means that the $k$th virtual node needs to be remitted to the physical node of another candidate domain; if the value of $v_{i}^{k}$ is 1, it means that the mapping scheme does not need to be changed.

(4) Optimal position of the particle itself $X_{sp}$: If the position of a particle can make the fitness function of the map lowest during the whole position change process, then this position is the optimal position of the particle itself.

(5) Global optimal position of particles $X_{gp}$: If the positions of all particles can make the fitness function of the map lowest during the whole change process, then this position is the global optimal position.

(6) a, b and c are constants, and the constraint relationship between them is a+b+c=1, which can be used to adjust the influence of three terms in the formula on velocity change.

(7) $\ominus$: When the two items are the same, take 1; when the two items are different, take 0.

(8) $\oplus$: After adding by bits, carry out rounding operation.

(9) $\otimes$: Decides whether to re-select the candidate node.

After PSO algorithm, an optimal VNE scheme is formed. During the generation of the mapping scheme, the node mapping scheme is first generated, that is, the corresponding physical node of each virtual node. After the node mapping scheme is formed, the link mapping scheme is formed according to whether there is a virtual link between two nodes in the pseudo-topology. The link mapping scheme is generated by Floyd algorithm. After the whole mapping scheme is formed, the global controller generates the corresponding request for each physical domain and sends it down to the local controller. Finally, the local controller maps each physical domain.

\subsection{Substrate Network Mapping}
The substrate network mapping is the last step in the mapping process, when the substrate network receives the mapping request issued by the global controller, it begins to map the physical domain. Since the local controller at the beginning passes the pseudo topology to the global controller, so the mapping scheme generated by the global controller is based on the pseudo topology. When the global controller sends the mapping result down to the local controller, the local controller can only determine the link between the two nodes, but there is the case that the link is connected across nodes, therefore, the shortest path between two nodes needs to be implemented by Floyd shortest path algorithm.

\section{Implementation of Algorithm}
In this section, we introduce the key steps in the algorithm implementation process.

\subsection{Candidate Physical Nodes Selection Algorithm}
The candidate physical nodes selection algorithm is the key of our scheme. In order to better meet users' demand for low delay, we proposed a candidate node selection algorithm based on delay estimation. In this algorithm, we use formula (5) to measure the delay of the physical nodes, and choose the physical nodes with the minimum delay as the candidate physical nodes of the virtual nodes.
\begin{algorithm}[h]
\caption{Candidate Physical Node Selection Algorithm}
\label{alg:Framwork}
\begin{algorithmic}[1]
    \Require
    Virtual network request $G_{V}=\left \{N_{V},L_{V} \right \}$; Substrate network topology $G_{S}=\left \{N_{S},L_{S} \right \}$;
    \Ensure
    Candidate set of physical nodes;
    \For {each physical node}
     \State compute the delay of each physical node;
    \EndFor
    \State sort the nodes by delay;
    \For {each substrate node}
     \State find the least costly node in turn;
     \If {$marked(N_{S})$ is false}
      \State mark the physical node as a candidate node;
      \EndIf
     \EndFor
    \State\Return the candidate physical nodes;
  \end{algorithmic}
\end{algorithm}

\subsection{PSO Algorithm}
Particle swarm optimization (PSO) algorithm is used to optimize the pseudo-topology received by the global controller, aiming to find the optimal solution of virtual network mapping through a limited number of iterations. In the process of PSO, the position and velocity of the nodes are firstly initialized, and then the particle velocity is calculated and the particle position is changed within the number of iterations. In this paper, we calculate the mapping of the time delay to measure whether the position and velocity  of particles need to be changed.
\begin{algorithm}[h]
\caption{The PSO Algorithm}
\label{alg:Framwork}
\begin{algorithmic}[1]
    \Require
    Virtual network request $G_{V}=\left \{N_{V},L_{V} \right \}$; Substrate network pseudo-topology;
    \Ensure
    Node mapping scheme;
    \For{iteration\_time=0 to N}
    \For {each particle}
    \State calculate the velocity of each particle;
    \State change the position of each particle;
    \State randomly select whether to reset the position of each particle;
    \State calculate the mapping delay of each particle;
      \If {need to reset the reposition the particle}
      \State reposition the particle;
      \EndIf
      \If {the delay of the particle in its current position $<$ the delay of a particle in an optimal position}
      \State update the historical optimal position of the particle;
      \EndIf
      \If {the delay of the particle in its current position $<$ the delay of a particle in an global optimal position}
      \State update the global optimal position of the particle;
      \EndIf
    \EndFor
    \EndFor
    \State \Return the global optimal position of particles;
  \end{algorithmic}
\end{algorithm}

\subsection{Virtual Network Pre-mapping Algorithm}
After the global controller finds the optimal mapping scheme through the PSO algorithm, it first sends a mapping request to the candidate physical nodes corresponding to the virtual node, and then sends a mapping request to the physical link.
\begin{algorithm}[h]
\caption{Virtual Network Pre-mapping Algorithm}
\label{alg:Framwork}
\begin{algorithmic}[1]
    \Require
    Candidate nodes; Substrate network pseudo-topology;
    \Ensure
    Virtual network embedding results;
    \State initializes the mapping request;
     \For {each candidate node}
     \State add the candidate physical node to the mapping request;
     \EndFor
      \If {link exist between virtual node1 and virtual node2}
      \State add the virtual link to the mapping request;
      \EndIf
    \State \Return mapping result;
  \end{algorithmic}
\end{algorithm}

\subsection{Substrate Network Mapping Algorithm}
After the mapping result is delivered to the local controller, the physical domain managed by the local controller is mapped. Firstly, the physical node is mapped, and then the physical link is mapped. Finally, the delay of the VNE is calculated, that is, the objective function defined in formula (5). The smaller the result, the more successful the mapping is.
\begin{algorithm}[h]
\caption{Substrate Network Mapping Algorithm}
\label{alg:Framwork}
\begin{algorithmic}[1]
    \Require
    Mapping result, Substrate network topology;
    \Ensure
    Delay of the system;
    \State initializes the mapping request;
     \For {each candidate node}
     \State map the physical nodes;
     \State calculate the delay of the physical node;
     \EndFor
      \If {links exist between physical node1 and physical node2}
      \State map the physical link;
      \State calculate the delay of physical link;
      \EndIf
    \State \Return the delay of system;
  \end{algorithmic}
\end{algorithm}

\subsection{Time Complexity Analysis}
In this section, we denote the number of virtual nodes and substrate nodes by $\left | N_{V} \right |$ ,$\left | N_{S} \right |$ respectively, and denote the number of candidate substrate nodes by $\left | N_{C} \right |$. In PSO  algorithm, the number of iteration is denoted by $\left |iteration \right |$, and the number of particles is represented by $\left |particle \right |$. The time complexity of algorithm 1 is $O\left ( \left | N_{S} \right |\right )$, the time complexity of algorithm 2 is $O\left ( \left |iteration \right |\left |particle \right |\right )$, the time complexity of algorithm 3 is $O\left ( \left | N_{C} \right |+\left | N_{S} \right |^{2}\right )$, and the time complexity of algorithm 4 is O( $\left | N_{C} \right |+\left | N_{C} \right |^{2}$ ).

\section{Simulation Experiment and Analysis}
This section is divided into two parts. In the first part, we describe the setting of simulation environment and experimental parameters. In the second part, we conduct simulation experiments and give simulation results.

\subsection{Experimental Environment Settings}
The environment used in this paper is 4GB memory, 64-bit windows10 operating system. The simulation code is written in the java language on eclipse, the network topology used in the simulation experiment was randomly generated by eclipse, the final comparison of the test results was drawn using Origin. The parameters of the simulation experiment are shown in TABLE \ref{tbl-eg1}:

\begin{table}
\begin{tabular}{@{\hspace{1mm}}l@{\hspace{1mm}}lll@{\hspace{1mm}}l@{\hspace{1mm}}}\hline
Parameter Items & The Range\\\hline
the number of substrate domains      & 5            \\
the number of substrate nodes        & 30          \\
substrate node resource              & U[100,300]   \\
substrate link resources             & U[1000,3000] \\
substrate link delay                 & U[1,10]      \\
intra-domain link connection rate    & 0.6      \\ \hline
the number of virtual nodes          & 4     \\
virtual node CPU capacity            & U[1,10]      \\
virtual link bandwidth               & U[1,10]      \\ \hline
the number of particles in PSO       & 10      \\
the number of iterations in PSO      & 50      \\
the value of a in PSO                & 0.3      \\
the value of b in PSO                & 0.4     \\
the value of c in PSO                & 0.3      \\
the mutation probability in PSO      & 0.1      \\ \hline
\end{tabular}
\caption{The Settings of Parameters.\\}
\label{tbl-eg1}
\end{table}

\subsection{Experimental Results and Analysis}
In order to verify the performance of the DP-VNE algorithm proposed in this paper, we designed four simulation experiments. The algorithms compared with DP-VNE include the MP-VNE algorithm, the VNE-PSO algorithm proposed in \cite{gengruiwen} and the MC-VNE algorithm proposed in \cite{luhancheng}, they were compared in terms of delay and mapping success rate. TABLE \ref{tbl-eg2} compares the ideas of the four algorithms.

\begin{table}
\begin{center}
\begin{tabular}{|@{\hspace{1mm}}l@{\hspace{1mm}}|l|ll@{\hspace{1mm}}|l@{\hspace{1mm}}|}\hline
Notation & The idea of algorithm \\
\hline
 DP-VNE&Firstly, the candidate physical node with the minimum delay \\
       &is selected according to the estimated delay unit, and then \\
       &the PSO algorithm is used to optimize the mapping process, \\
       &and finally the optimal mapping scheme is obtained.  \\\hline
 MP-VNE&The multi-objective optimization problem is transformed \\
       &into a single-objective optimization problem by weight, the \\
       &boundary nodes are selected by the mapping cost formula. \\\hline
VNE-PSO&Firstly, the candidate physical nodes are selected from the \\
       &nodes closest to the boundary nodes, and then PSO algorithm \\
       &is used to solve the node mapping scheme. \\\hline
 MC-VNE&The kruskal algorithm is used to select the link with the \\
       &lowest link unit price for mapping, and the node mapping \\
       &scheme is determined by the link mapping scheme. \\\hline
\end{tabular}
\end{center}
\caption{The compared three methods.}
\label{tbl-eg2}
\end{table}

Next, we designed four experiments to compare the performance of the four algorithms. The four experiments compared the above algorithms from four aspects of average link delay, average node delay, average overall delay and mapping success rate respectively. Obviously, while maintaining the stability of mapping success rate, the DP-VNE algorithm also has a significant advantage in the aspect of time delay.

\textbf{Experiment 1:} The average link delay performance comparisons.

\begin{figure}
\begin{center}
\scalebox{0.3}{\includegraphics{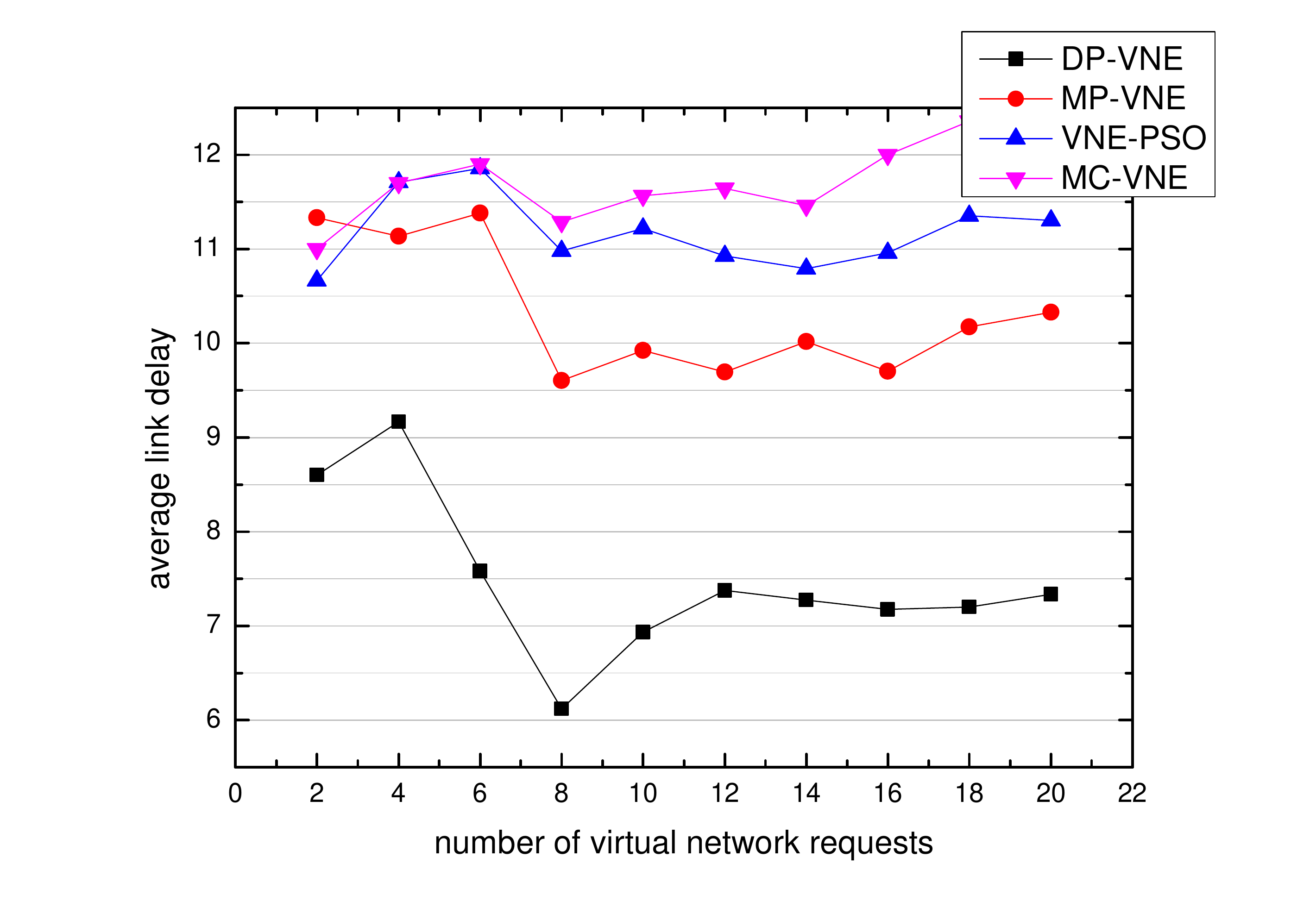}}
\end{center}
\caption{The average link delay performance comparisons.}
\label{fig-eg4}
\end{figure}

This experiment compares the average link delay of DP-VNE with MP-VNE, VNE-PSO and MC-VNE under different virtual network request numbers. As can be seen from FIGURE \ref{fig-eg4}, with the increase of the number of virtual network requests, the average link delay of the four algorithms generally tends to be stable. The picture shows that when the number of virtual network requests is less than 8, the average delay of links is higher; when the number of virtual network requests is more than 8, the average delay of links tends to be flat. The average link delay of the proposed DP-VNE algorithm is significantly lower than that of MP-VNE, VNE-PSO and MC-VNE algorithms, while the average link delay of MP-VNE algorithm is also slightly lower than that of VNE-PSO algorithm.

\textbf{Experiment 2:} The average node delay performance comparisons.

\begin{figure}
\begin{center}
\scalebox{0.3}{\includegraphics{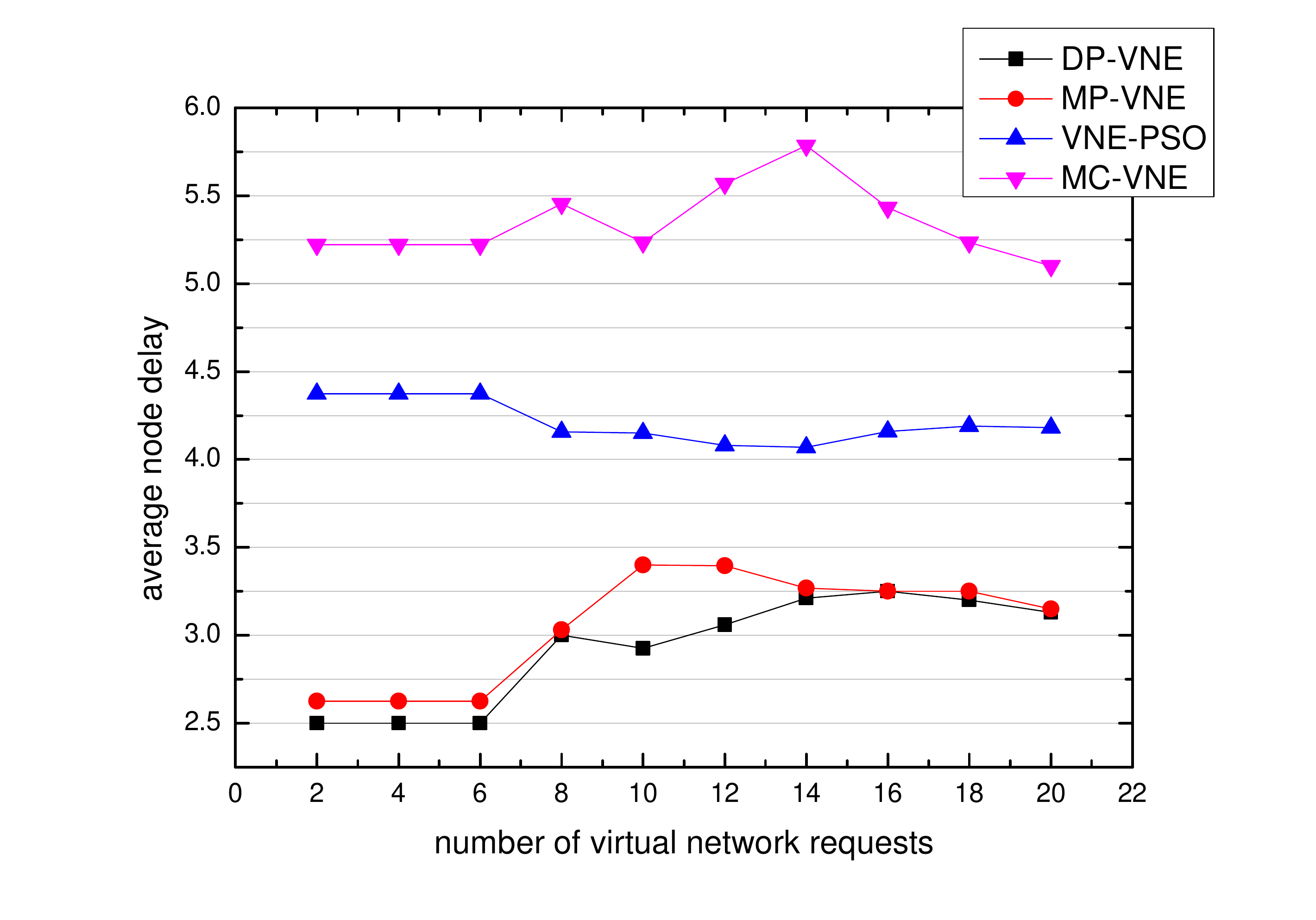}}
\end{center}
\caption{The average node delay performance comparisons.}
\label{fig-eg5}
\end{figure}

This experiment compares the changes of the average node delay of these four algorithms under different numbers of virtual network requests. It can be clearly seen from FIGURE \ref{fig-eg5} that with the increase of the number of virtual network requests, the average node delay of the four algorithms shows a trend of slowly increasing. When the number of virtual network requests is less than 6, the node average delay is constant, because when the number of virtual network requests is small, the substrate resources are sufficient, so the shortest delay can be guaranteed for mapping. The average node delay of the proposed DP-VNE algorithm is slightly lower than that of the MP-VNE algorithm and significantly lower than that of the VNE-PSO algorithm. As can be seen from the comparison diagram, the DP-VNE algorithm proposed in this paper is obviously optimized in terms of node delay.

\textbf{Experiment 3:} The average overall delay performance comparisons.

\begin{figure}
\begin{center}
\scalebox{0.3}{\includegraphics{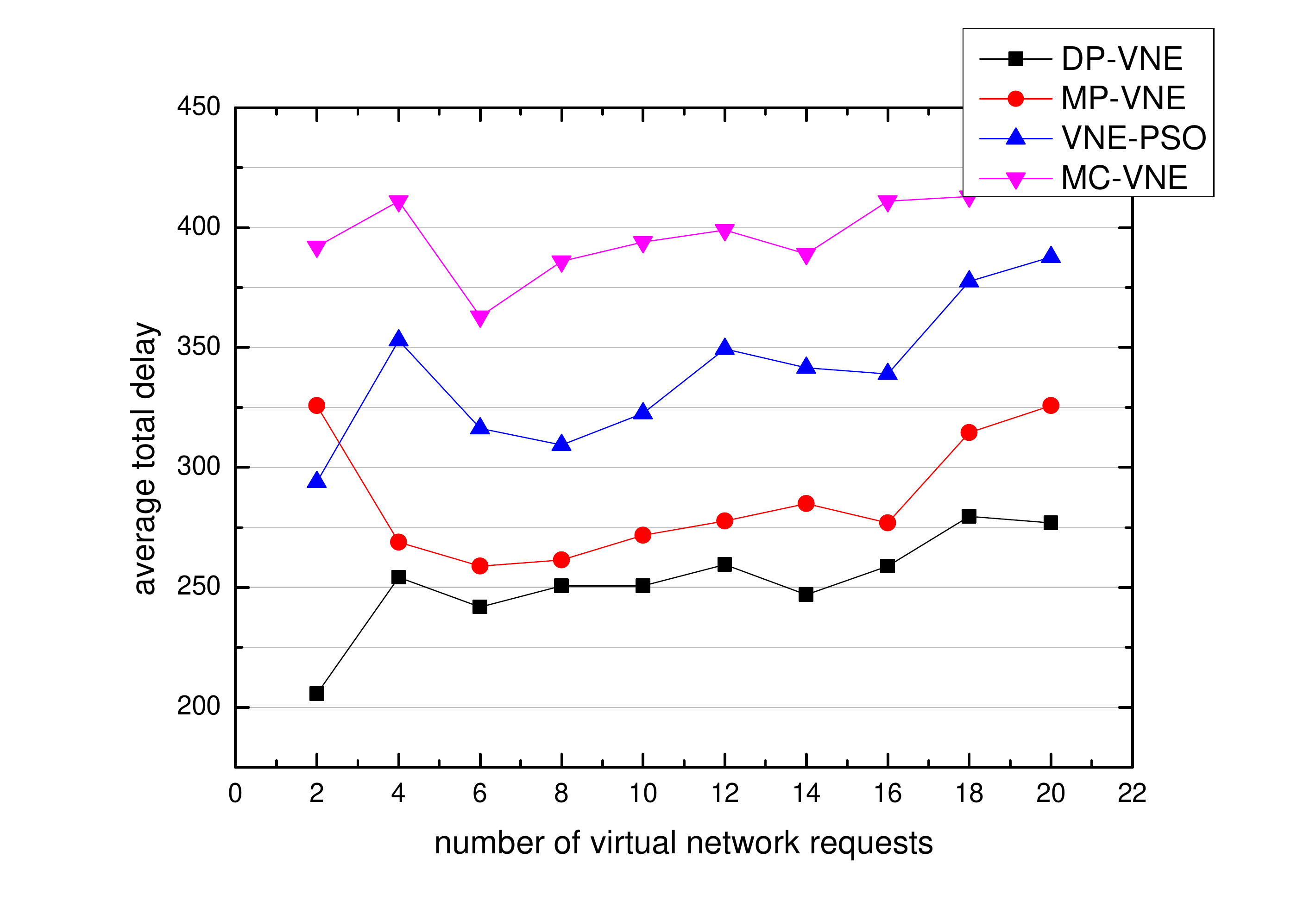}}
\end{center}
\caption{The average total delay performance comparisons.}
\label{fig-eg6}
\end{figure}

This experiment compares the total delay of the four algorithms under different numbers of virtual network requests, which can best represent the delay performance because it combines the node delay and link delay of each algorithm. As can be seen from FIGURE \ref{fig-eg6}, with the increase of the number of virtual network requests, the total delay fluctuation range of the four algorithms is gentle and generally tends to be stable. The total delay of the DP-VNE algorithm proposed in this paper is significantly lower than that of the other four algorithms. In this experiment, the total delay of the VNE-PSO algorithm is lower than that of the MP-VNE algorithm, and the average total delay of MC-VNE is the highest, indicating that the heuristic algorithm is still superior to the simple greedy algorithm in terms of delay. To sum up, the DP-VNE algorithm proposed in this paper compared with the other three algorithms, significantly decreases the delay attribute in the process of virtual network mapping and achieves the goal of optimizing the delay.

\textbf{Experiment 4:} The VNE's success rate comparisons.

\begin{figure}
\begin{center}
\scalebox{0.3}{\includegraphics{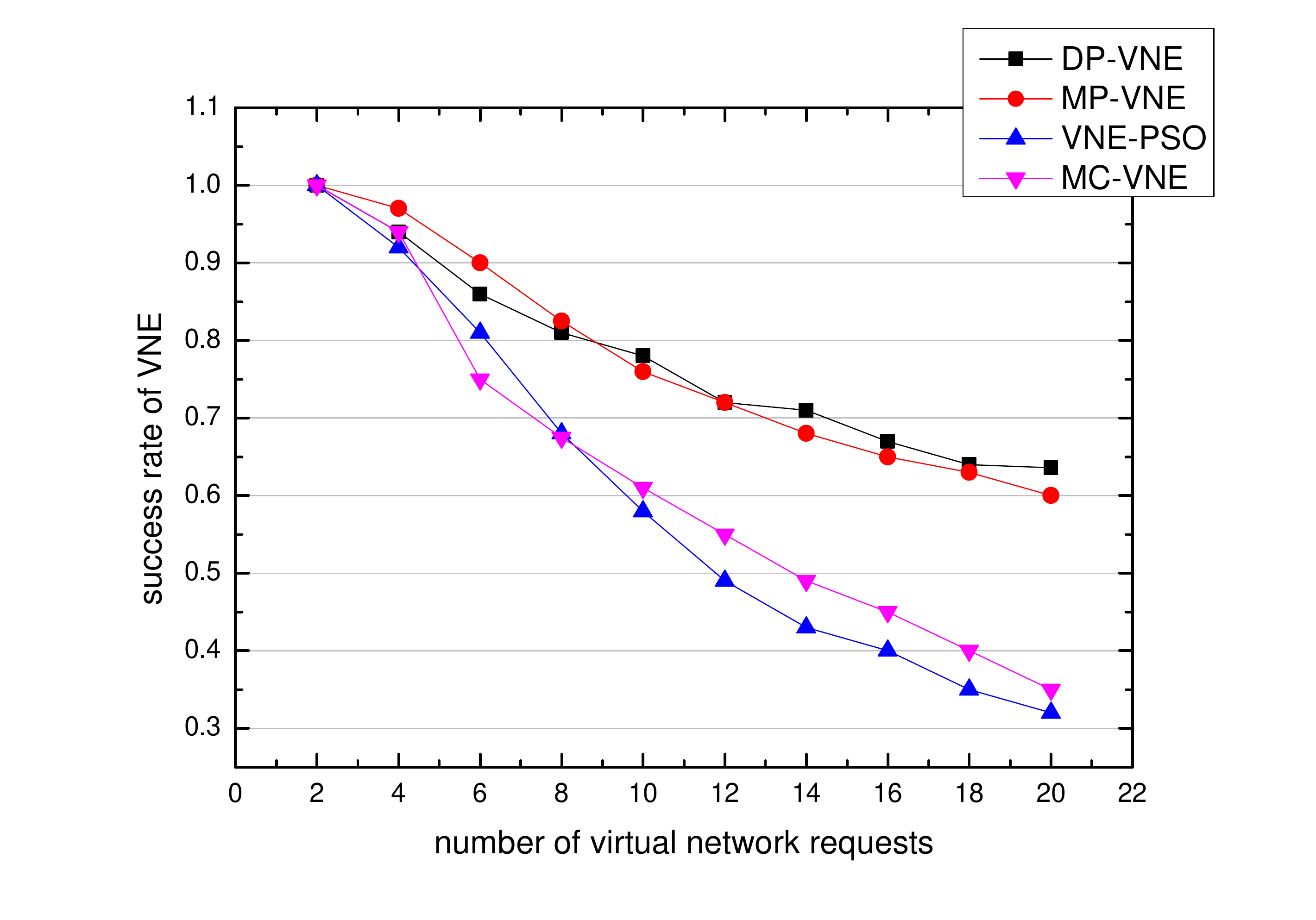}}
\end{center}
\caption{The VNE's success rate comparisons.}
\label{fig-eg7}
\end{figure}

The mapping success rate of the four algorithms is shown in FIGURE \ref{fig-eg7}. However, the virtual request acceptance rate of DP-VNE and MP-VNE is significantly better than that of VNE-PSO and MC-VNE. The DP-VNE algorithm proposed in this paper is slightly better than that of MP-VNE and has the highest mapping success rate among the four algorithms. It can be seen that DP-VNE and MP-VNE have a steady trend of mapping success rate around 0.6 over scale, while the other two algorithms still maintain a downward trend at 0.3.

Simulation experiments show that, this algorithm has outstanding contribution in reducing delay, which mainly depends on the following points:

(1) The DP-VNE algorithm proposed in this paper uses the layered SDN virtual network mapping architecture, which achieves the separation of control plane and data plane compared with the traditional virtual network mapping architecture, and can effectively improve the mapping efficiency.

(2) In the selection of candidate physical nodes, a delay prediction formula is defined in this algorithm. Through this formula, the physical node with the smallest delay measurement can be selected as the candidate node. Therefore, in experiment 2, the average delay of nodes of this algorithm is much lower than that of the other two algorithms.

(3) When the algorithm uses particle swarm optimization to optimize the mapping scheme, it defines an objective function, which sets the weight of delay to the maximum and integrates the delay attributes of nodes and links. Therefore, when iteratively selecting the optimal mapping scheme, the delay is first considered.

(4) Floyd algorithm is always used in the link mapping in this paper from selection to mapping. This algorithm ensures that when the node mapping result is determined, an shortest path can be selected between the two physical nodes, thus, the link delay is further reduced.

\section{Conclusion}
Network virtualization is an advanced virtualization technology, which solves the ossification problem of traditional networks to a large extent. With the emergence of more and more delay-sensitive applications, users have a large demand for low delay. Based on this demand, this paper proposed a multi-domain virtual network embedding algorithm based on delay prediction (DP-VNE), which defines a DelayUnit to select candidate physical nodes for virtual nodes, and uses PSO algorithm to generate the optimal mapping results, the whole process is based on multi-domain SDN architecture. The simulation results show that the DP-VNE algorithm has an outstanding contribution in reducing the system delay while keeping other indexes unaffected. In the future, we will study how to improve the performance of other aspects while ensuring the minimum delay.

\section{Acknowledgments}
This work is supported by ``the Fundamental Research Funds for the Central Universities'' of China University of Petroleum (East China) (Grant No. 18CX02139A). The authors also gratefully acknowledge the helpful comments and suggestions of the reviewers, which have improved the presentation.

\appendix

\end{document}